\documentstyle[12pt,a4,epsf]{article}

\makeatletter
\@addtoreset{equation}{section}
\makeatother

\newcommand{\VEV}[1]{\left\langle #1 \right\rangle}

\newcommand{\nn}{\nonumber}

\newcommand{\bequ}{\begin{equation}}
\newcommand{\eequ}{\end{equation}}
\newcommand{\beqn}{\begin{eqnarray}}
\newcommand{\eeqn}{\end{eqnarray}}

\begin{document}
\begin{titlepage}

\begin{flushright}
hep-ph/0206048\\
KUNS-1789\\

\today
\end{flushright}

\vspace{4ex}

\begin{center}
{\large \bf
The GUT? Neutrino bi-large mixing and proton decay
\footnote
{Talk presented at NOON2001 held at ICRR, Kashiwa(Japan),
Dec. 5-8 2001.
}
}

\vspace{6ex}

\renewcommand{\thefootnote}{\alph{footnote}}
Nobuhiro Maekawa\footnote
{e-mail: maekawa@gauge.scphys.kyoto-u.ac.jp
}

\vspace{4ex}
{\it Department of Physics, Kyoto University, Kyoto 606-8502, Japan}
\end{center}

\renewcommand{\thefootnote}{\arabic{footnote}}
\setcounter{footnote}{0}
\vspace{6ex}

%--------------------<<   abstract   >>--------------------
\begin{abstract}
In this talk, we introduce a new scenario of grand unified theory
(GUT) with anomalous $U(1)_A$ gauge symmetry, which can explain 
doublet-triplet splitting, quark and lepton
masses and mixing angles. In neutrino sector, the scenario
realizes LMA solution for solar neutrino problem and large 
$U_{e3}=O(0.1)$. Moreover, the scenario predicts
that the main decay mode of proton is from dimension 6 operators and
the lifetime of proton must be near the present limit.
The realization of gauge coupling unification requires
that the cutoff scale of the scenario must be around the usual GUT
scale $\Lambda_G\sim 2\times 10^{16}$ GeV, which is smaller than
the Planck scale. It may suggest the extra dimension in which gauge
fields in visible sector do not propagate.
This talk is based on the papers.\cite{maekawa,maekawa2,maekawa3,BM,MY}

\end{abstract}

\end{titlepage}

%--------------------<<   section    >>--------------------
\section{Introduction}
The most of people regard the standard model as the real theory which
describe our world but does not satisfy the model as the final theory,
because there are a lot of things which are not explained by the model;
unstability of the weak scale due to quadratic divergent loop correction
to the Higgs mass term: the miracle anomaly cancellation between quark and
leptons: the origin of hierarchies of gauge and Yukawa couplings: the origin
of small mixings in quark sector and large mixings in lepton sector: the
charge quantization: no gravity, etc.
The idea of grand unified theories (GUT)\cite{georgi} not only
explain the hierarchy of three gauge couplings in the standard model,
anomaly cancellation and charge quantization, but also 
gives a natural unification of quark and leptons in a few multiplets in
a simple gauge group. 
Since supersymmetry (SUSY) can stabilize the weak scale, SUSY GUT is one
of the most promising model beyond the standard model. 
Unfortunately, it is not so easy to obtain the realistic SUSY GUT, 
because it is difficult to solve the doublet-triplet
splitting problem
\cite{DTsplitting,DW,BarrRaby}
 with stable proton and to obtain realistic quark and lepton 
mass matrices. 
On the other hand, it is known
\cite{Ibanez,Ramond,Dreiner} that the hierarchy of Yukawa couplings in
the supersymmetric standard model 
are realized by introducing anomalous $U(1)$ gauge symmetry,
\cite{U(1)} whose anomaly is cancelled by the Green-Schwarz mechanism.
\cite{GS}
Of course, it is not so straightforward to extend the argument in the SUSY
standard model into in the GUT scenario, especially with large neutrino 
mixing angles, but it is important to examine the GUT scenario with anomalous
$U(1)_A$ gauge symmetry.\cite{bando,IKNY,shafi}

Recently, in a series of papers,\cite{maekawa,maekawa2,maekawa3,BM,MY}
 the interesting GUT scenario with
anomalous $U(1)_A$ gauge symmetry has been proposed with $SO(10)$ unified
group\cite{maekawa} and with $E_6$ unified group.\cite{BM,MY}
In the scenario,
the anomalous $U(1)_A$ gauge symmetry plays an important role not only
in obtaining realistic quark and lepton mass matrices, including
bi-large neutrino mixings (LMA for solar neutrino problem) but also 
in solving doublet-triplet splitting problem. 
Moreover, since generic interactions are allowed to be introduced,
it predicts the mass spectrum of superheavy fields and (GUT) symmetry
breaking scales once the symmetry of the theory is fixed. It is surprising
that the success of coupling unification in the minimal SUSY standard model
(MSSM) can be naturally explained in the scenario, though the mass spectrum 
of superheavy fields does not respect $SU(5)$ symmetry.
\cite{maekawa3}
It is shown that the gauge coupling unification requires the cutoff scale
must be around the usual GUT scale $\Lambda_G=2\times 10^{16}$ GeV and 
the unification scale is just below the usual GUT scale if all the fields
except those of MSSM have superheavy masses.
\cite{maekawa3} It is interesting that this
result is independent of the detail of Higgs sector. The GUT with anomalous 
$U(1)_A$ gauge symmetry with a simple unified gauge group predicts the above
result. Therefore, proton decay via dimension 6 operators may be seen in near
future experiments. Moreover, once SUSY breaking parameters are introduced,
the $\mu$ problem is naturally solved \cite{maekawa2}
and natural suppression of flavor changing
neutral current (FCNC) is realized in $E_6$ GUT.\cite{BM}

As introduced in the above, the scenario predicts the proton decay
via dimension 6 operators even though low energy SUSY is required.
Moreover, bi-large neutrino mixings are obtained (especially,
LMA solution for the solar neutrino problem is predicted).
The scenario predicts large $U_{e3}\sim O(0.1)$ and small 
$\tan \beta$. 

It is interesting that all the above solutions are realized in non-trivial 
ways once only several anomalous $U(1)_A$ charges are determined.
Actually, the input parameters are only 8 integer anomalous $U(1)_A$ charges  
(+3 for singlet Higgs) for the Higgs sector and 3(or 4) (half) integer 
charges for the matter sector in $E_6$ (or $SO(10)$) GUT. 
In this talk, we will explain some of them.

\section{Doublet-triplet splitting}
One of the most interesting feature of anomalous $U(1)_A$ gauge theory is
that the vacuum expectation values (VEV) are determined by anomalous
$U(1)_A$ charges as
\begin{eqnarray}
\VEV{Z^+}&=& 0, \label{VEV+} \\
\VEV{Z^-}&\sim&  \lambda^{-z^-},\label{VEV-}
\end{eqnarray}
where $Z^\pm$ are singlet operators with the charges $z^+>0$ and $z^-<0$, 
and $\lambda=\VEV{\Theta}/\Lambda$. Here $\Theta$ is a Froggatt-Nielsen
field.
\cite{FN}
 Through this paper, we use unit in which the cutoff $\Lambda=1$
and denote all the superfields by uppercase letters and their anomalous
$U(1)_A$ charges by the corresponding lowercase letters.
Such VEVs do not change the order of the coefficients obtained by
the Froggatt-Nielsen mechanism:
\begin{equation}
W=\left(\frac{\Theta}{\Lambda}\right)^{x+y+z}XYZ\rightarrow \lambda^{x+y+z}XYZ,
\end{equation}
if the total charge $x+y+z$ of the operator $XYZ$ is positive.
Note that even if the operator $\frac{Z^-}{\Lambda}$ is used instead of 
$\left(\frac{\Theta}{\Lambda}\right)^{z^-}$ in the interactions, the order
of the coefficients does not change. This feature is critically different
from the naive expectation that the contribution from the higher dimensional 
operators is more suppressed. 
If the total charge $x+y+z$ is negative, such interaction is not allowed by
the anomalous $U(1)_A$ gauge symmetry because only negatively charged fields
have non-vanishing VEVs. This is called SUSY zero mechanism. 
Note that this mechanism leads to the finite number of non-renormalizable 
interactions, and therefore we can control the generic superpotential.

Actually, under the vacua (\ref{VEV+}), the generic 
superpotential to determine the VEVs of $Z^-$ can be written as
\begin{equation}
W=\sum_i^{n_+}W_{Z_i^+}, \label{W}
\end{equation}
where $W_X$ denotes the terms linear in the $X$ field.
This is because the $F$-flatness conditions of negatively charged fields
are automatically satisfied and the terms with more than two positively charged
fields do not contribute in the $F$-flatness condition of positively charged
fields. 

Let us discuss an $SO(10)$ GUT model with anomalous $U(1)_A$ gauge symmetry
in which doublet-triplet splitting is naturally realized. 
The Higgs content is 
listed in Table I. 
\begin{table}
\begin{center}
Table I. Typical values of anomalous $U(1)_A$ charges.\\
\vspace{1mm}
\begin{tabular}{|c|c|c|} 
\hline
                  &   non-vanishing VEV  & vanishing VEV \\
\hline 
{\bf 45}          &   $A(a=-1,-)$        & $A'(a'=3,-)$      \\
{\bf 16}          &   $C(c=-4,+)$        
                  & $C'(c'=3,-)$      \\
${\bf \overline{16}}$&$\bar C(\bar c=-1,+)$ 
                  & $\bar C'(\bar c'=6,-)$ \\
{\bf 10}          &   $H(h=-3,+)$        & $H'(h'=4,-)$      \\
{\bf 1}           &$\Theta(\theta=-1,+)$,$Z(z=-2,-)$,
                  $\bar Z(\bar z=-2,-)$& $S(s=5,+)$ \\
\hline
\end{tabular}
\end{center}
\end{table}
Here the symbols $\pm$ denote the $Z_2$ parity.
The VEVs of the negatively charged Higgs fields are determined by
the superpotential
\begin{equation}
W=W_{A'}+W_{C'}+W_{\bar C'}+W_{H'}+W_S.
\end{equation}
We do not have spaces enough to explain the vacuum structure in detail,
so we here point out only one good feature in realizing the doublet-triplet
splitting. 

If $-3a\leq a^\prime < -5a$,
the superpotential $W_{A^\prime}$ is in general
written 
\begin{equation}
W_{A^\prime}=\lambda^{a^\prime+a}\alpha A^\prime A+\lambda^{a^\prime+3a}(
\beta(A^\prime A)_{\bf 1}(A^2)_{\bf 1}
+\gamma(A^\prime A)_{\bf 54}(A^2)_{\bf 54}),
\end{equation}
where the suffices {\bf 1} and {\bf 54} indicate the representation 
of the composite
operators under the $SO(10)$ gauge symmetry, and $\alpha$, 
$\beta$ and $\gamma$ are parameters of order 1. Here we assume 
$a+a^\prime+c+\bar c<0$
to forbid the term $\bar C A^\prime A C$, which destabilizes the 
DW form of the VEV $\VEV{A}$. 
The $D$-flatness condition requires the VEV
$\VEV{A}=i\tau_2\times {\rm diag}(x_1,x_2,x_3,x_4,x_5)$, and the 
$F$-flatness conditions of the $A^\prime$ field requires
$x_i(-\alpha\lambda^{-2a}
+(2\beta-\frac{\gamma}{5})(\sum_j x_j^2)+\gamma x_i^2)=0$. 
This allows only two solutions, $x_i^2=0$ and 
$x_i^2=\frac{\alpha}{(1-\frac{N}{5})\gamma+2N\beta}\lambda^{-2a}\equiv v^2$. 
Here $N=0$ -- 5 is the number of $x_i=v$ solutions.
When $N=3$, the vacuum becomes
$\VEV{A({\bf 45})}_{B-L}=\tau_2\times {\rm diag}
(v,v,v,0,0)$, which breaks $SO(10)$ into
$SU(3)_C\times SU(2)_L\times SU(2)_R\times U(1)_{B-L}$
at the scale $\Lambda_A\equiv \VEV{A}\sim \lambda^{-a}$.
 This Dimopoulos-Wilczek form of the VEV plays an
important role in solving the DT splitting problem.
Actually through the interaction
$W=H'AH$, the DW type of the VEV gives superheavy masses only to the 
triplet Higgs, and therefore the doublet Higgs remains massless. 
Taking account of the mass term $H'^2$, only one pair of Higgs doublets
becomes massless. 

Note that the higher terms $A^\prime A^{2L+1}$ $(L>1)$ are 
forbidden by the SUSY zero mechanism. If they were allowed, 
the number of possible VEVs other than the DW form would 
become larger, and thus it would become less natural to obtain 
the DW form. 
This is a critical point of this mechanism, and the anomalous 
$U(1)_A$ gauge symmetry plays an essential role in forbidding 
the undesired terms.

The spinor Higgs fields 
$C$ and $\bar C$ break $SU(2)_R\times U(1)_{B-L}$ into $U(1)_Y$ 
by developing 
$\VEV{C}(=\VEV{\bar C}\equiv \Lambda_C\sim \lambda^{-(c+\bar c)/2}$).
Then this model becomes MSSM at a low energy scale.

\section{Quark and lepton mass matrices}
One of the most attractive features of grand unified theory is to 
unify the quark and lepton into fewer multiplets. For example, 
in $SO(10)$ GUT scenario, a ${\bf 16}$ representation field contains
one family quark and lepton fields including right-handed neutrino
field. However, this attractive feature directly leads to unrealistic
Yukawa relations. For example, if we introduce $3\times{\bf 16}$ 
$\Psi_i (i=1,2,3)$ for 3 family quark and leptons, the Yukawa couplings
are obtained from the interaction
\begin{equation}
W=Y_{ij}\Psi_i\Psi_j H
\end{equation}
as $Y_u=Y_d$ and $Y_d=Y_e,$
which lead to unrealistic mass relation.
We have to pick up the VEV $\VEV{C}$ in the Yukawa matrices to avoid
the former unrealistic relation $Y_u=Y_d$, and the VEV $\VEV{A}$
to avoid the latter unrealistic relation $Y_d=Y_e$.
In our scenario, we introduce an additional matter field $T({\bf 10})$.
Then after breaking the GUT gauge group into the standard model gauge
group, one pair of vector-like fields ${\bf 5}$ and ${\bf \bar 5}$ of
$SU(5)$ becomes massive. The mass matrix is 
obtained from the interaction
\begin{equation}
W=\lambda^{\psi_i+t+c}\Psi_iTC+\lambda^{2t}T^2
\end{equation}
as 
\begin{equation}
{\bf 5}_T ( \lambda^{t+\psi_1+(c-\bar c)/2}, 
               \lambda^{t+\psi_2+(c-\bar c)/2}, 
               \lambda^{t+\psi_3+(c-\bar c)/2}, \lambda^{2t})
\left(
\begin{array}{c}  {\bf \bar 5}_{\Psi1} \\ {\bf \bar 5}_{\Psi2} \\ 
{\bf \bar 5}_{\Psi3}
 \\ {\bf \bar 5_T}
\end{array}
\right),
\end{equation}
where actually the VEV $\VEV{\bar C}=\VEV{C}\sim \lambda^{-(c+\bar c)/2}$
appear in the mass matrix.
Since $\psi_3<\psi_2<\psi_1$, 
the massive mode ${\bf \bar 5}_M$, the partner of ${\bf 5}_T$, must
be either 
${\bf \bar 5}_{\Psi3}(\Delta\equiv 2t-(t+\psi_3+(c-\bar c)/2)>0)$ or
${\bf \bar 5}_T(\Delta<0)$. 
The former case is interesting, and in this case,
the three massless modes 
$({\bf \bar 5}_1, {\bf \bar 5}_2, {\bf \bar 5}_3) $ 
can be written 
$({\bf \bar 5}_{\Psi1}+\lambda^{\psi_1-\psi_3}{\bf \bar 5}_{\Psi3}, 
{\bf \bar 5}_T+ \lambda^{\Delta}{\bf \bar 5}_{\Psi3},
{\bf \bar 5}_{\Psi2}+\lambda^{\psi_2-\psi_3}{\bf \bar 5}_{\Psi3})$. 
If we adopt their charges $(\psi_1,\psi_2,\psi_3,t)=(9/2,7/2,3/2,5/2)$
in addition to the charges of Higgs fields, then we can estimate quark 
and lepton mass matrices as
\begin{equation}
M_u=\left(
\begin{array}{ccc}
\lambda^6 & \lambda^5 & \lambda^3 \\
\lambda^5 & \lambda^4 & \lambda^2 \\
\lambda^3 & \lambda^2 & 1  
\end{array}
\right)\VEV{H_u},\quad
M_d=M_e=\lambda^2\left(
\begin{array}{ccc}
\lambda^4 & \lambda^{3.5} & \lambda^3 \\
\lambda^3 & \lambda^{2.5} & \lambda^2 \\
\lambda^1 & \lambda^{0.5} & 1
\end{array}
\right)\VEV{H_d}.
\label{quark}
\end{equation}
And for the neutrino sector, we take into account the interaction
\begin{equation}
\lambda^{\psi_i+\psi_j+2\bar c}\Psi_i\Psi_j\bar C\bar C,
\end{equation}
which lead to the right-handed neutrino masses
\begin{equation}
M_R=\lambda^{\psi_i+\psi_j+2\bar c}\VEV{\bar C}^2
=\lambda^{2n+\bar c-c}\left(
\begin{array}{ccc}
\lambda^6 & \lambda^5 & \lambda^3 \\
\lambda^5 & \lambda^4 & \lambda^2 \\
\lambda^3   & \lambda^2         & 1
\end{array}
\right).
\end{equation}
Since the Dirac neutrino mass is given by
\begin{equation}
M_{\nu_D}=\lambda^2\left(
\begin{array}{ccc}
\lambda^4 & \lambda^3 & \lambda \\
\lambda^{3.5} & \lambda^{2.5} & \lambda^{0.5} \\
\lambda^3   & \lambda^2         & 1 
\end{array}
\right)\VEV{H_u}\eta,
\end{equation}
the neutrino mass matrix is obtained by the seesaw mechanism
as
\begin{equation}
M_\nu=M_{\nu_D}M_R^{-1}M_{\nu_D}^T=\lambda^{4-2n+c-\bar c}\left(
\begin{array}{ccc}
\lambda^2          & \lambda^{1.5}  & \lambda \\
\lambda^{1.5} & \lambda & \lambda^{0.5} \\
\lambda            & \lambda^{0.5}  & 1 
\end{array}
\right)\VEV{H_u}^2\eta^2.
\end{equation}
Note that the ratio $\frac{m_{\nu_\mu}}{m_{\nu_\tau}}\sim \lambda$
is realized, that predicts LMA solution for the solar neutrino problem. 
It is interesting that we obtain the small mixing angles for
the Cabibbo-Kobayashi-Maskawa matrix
\begin{equation}
U_{\rm CKM}=
\left(
\begin{array}{ccc}
1 & \lambda &  \lambda^3 \\
\lambda & 1 & \lambda^2 \\
\lambda^3 & \lambda^2 & 1
\end{array}
\right),
\label{CKM}
\end{equation}
and the large mixing angles for the Maki-Nakagawa-Sakata matrix
\begin{equation}
U_{\rm MNS}=
\left(
\begin{array}{ccc}
1 & \lambda^{0.5} &  \lambda \\
\lambda^{0.5} & 1 & \lambda^{0.5} \\
\lambda & \lambda^{0.5} & 1
\end{array}
\right). 
\end{equation}
Since we use a rule that 
$\lambda^{0.5}+\lambda^{0.5}\sim \lambda^{0.5}$ in calculating the MNS
matrix and $\lambda^{0.5}\sim 0.5$, this model gives large mixing angles
\cite{SK,SNO}
for the atmospheric neutrino problem and for the solar neutrino problem.
And $U_{e3}\sim \lambda$ is predicted, which is around the present upper
limit given by CHOOZ.\cite{CHOOZ}
At this stage, the unrealistic GUT relation $Y_d=Y_e$ still remains.
However, in our scenario, the same amount of the Yukawa couplings are
given by the higher dimensional interactions
\begin{equation}
W=\lambda^{\psi_i+\psi_j+n a+h}\Psi_iA^n\Psi_jH
\end{equation}
by developing the VEV $\VEV{A}\sim \lambda^{-a}$.
It is critical that the Yukawa couplings from the higher dimensional 
interactions have not kept the unrealistic GUT relation.
Usually, the corrections from such higher dimensional interactions are
suppressed by the factor $\frac{\VEV{A}}{\Lambda}$. But in our scenario,
the suppression factor $\frac{\VEV{A}}{\Lambda}$ is just cancelled by the
enhancement factor $\lambda^{a}$ in the coefficients, and therefore
we can obtain the same order coefficients as from the tree interaction.
This is an attractive feature in our scenario, and 
the realistic mass matrices are naturally obtained.

\section{Gauge coupling unification}
First, we show the relation of the determinants of the mass spectrum of 
superheavy fields in terms of their anomalous $U(1)_A$ charges. 
If we use the notation of the fields $Q({\bf 3,2})_{\frac{1}{6}}$,
$U^c({\bf \bar 3,1})_{-\frac{2}{3}}$, $D^c({\bf \bar 3,1})_{\frac{1}{3}}$,
$L({\bf 1,2})_{-\frac{1}{2}}$, $E^c({\bf 1,1})_1,N^c({\bf 1,1})_0$,
$X({\bf 3,2})_{-\frac{5}{6}}$ and their conjugate fields, and
$G({\bf 8,1})_0$ and $W({\bf 1,3})_0$ with the standard gauge symmetry,
under $SO(10)\supset SU(5) \supset SU(3)_C\times SU(2)_L\times U(1)_Y$,
the spinor ${\bf 16}$, vector ${\bf 10}$ and the adjoint ${\bf 45}$
of $SO(10)$
are divided as
\begin{eqnarray}
{\bf 16}&\rightarrow &
\underbrace{[Q+U^c+E^c]}_{\bf 10}+\underbrace{[D^c+L]}_{\bf \bar 5}
+\underbrace{N^c}_{\bf 1},\\
{\bf 10}&\rightarrow &
\underbrace{[D^c+L]}_{\bf \bar 5}+\underbrace{[\bar D^c+\bar L]}_{\bf 5},\\
{\bf 45}&\rightarrow &
\underbrace{[G+W+X+\bar X+N^c]}_{\bf 24}
+\underbrace{[Q+U^c+E^c]}_{\bf 10}
+\underbrace{[\bar Q+\bar U^c+\bar E^c]}_{\bf \overline{10}}
+\underbrace{N^c}_{\bf 1}.
\end{eqnarray}
Then the determinants of the mass matrices $\bar M_I$ of superheavy 
fields $I=Q$,$U^c$,$E^c$,$D^c$,$L$,$G$,$W$,$X$ are estimated as
\begin{equation}
\det \bar M_I = \lambda^{\mbox{$\sum_i c_i$}},
\end{equation}
where $c_i$ are anomalous $U(1)_A$ charges of superheavy fields.

Secondly, 
the conditions of the gauge coupling unification
in using one loop renormalization group equations
\begin{equation}
\alpha_3(\Lambda_A)=\alpha_2(\Lambda_A)=
\frac{5}{3}\alpha_Y(\Lambda_A)\equiv\alpha_1(\Lambda_A),
\end{equation}
where 
$\alpha_1^{-1}(\mu>\Lambda_C)\equiv 
\frac{3}{5}\alpha_R^{-1}(\mu>\Lambda_C)
+\frac{2}{5}\alpha_{B-L}^{-1}(\mu>\Lambda_C)$,
are rewritten by the determinants of the mass matrices of
the superheavy fields. 
Here $\alpha_X=\frac{g_X^2}{4\pi}$, and 
the parameters $g_X (X=3,2,R,B-L,Y)$ are the gauge couplings of 
$SU(3)_C$, $SU(2)_L$, $SU(2)_R$, $U(1)_{B-L}$ and $U(1)_Y$, 
respectively.

The gauge couplings at the scale $\Lambda_A$ are roughly given by
\begin{eqnarray}
\alpha_1^{-1}(\Lambda_A)&=&\alpha_1^{-1}(M_{SB})
+\frac{1}{2\pi}\left(b_1\ln \left(\frac{M_{SB}}{\Lambda_A}\right)
+\sum_i \Delta b_{1i}\ln \left(\frac{m_i}{\Lambda_A}\right)
-\frac{12}{5}\ln \left(\frac{\Lambda_C}{\Lambda_A}\right)\right), 
\label{alpha1} \nn \\
&&  \\
\alpha_2^{-1}(\Lambda_A)&=&\alpha_2^{-1}(M_{SB})
+\frac{1}{2\pi}\left(b_2\ln \left(\frac{M_{SB}}{\Lambda_A}\right)
+\sum_i \Delta b_{2i}\ln \left(\frac{m_i}{\Lambda_A}\right)\right), \\
\alpha_3^{-1}(\Lambda_A)&=&\alpha_3^{-1}(M_{SB})
+\frac{1}{2\pi}\left(b_3\ln \left(\frac{M_{SB}}{\Lambda_A}\right)
+\sum_i \Delta b_{3i}\ln \left(\frac{m_i}{\Lambda_A}\right)\right), 
\end{eqnarray}
where $M_{SB}$ is a SUSY breaking scale, 
$(b_1,b_2,b_3)=(33/5,1,-3)$ are the 
renormalization group coefficients
for the minimal SUSY standard model (MSSM),
and $\Delta b_{ai}\ (a=1,2,3)$ are the corrections to the coefficients 
from the massive fields with mass $m_i$.
The last term in Eq. (\ref{alpha1}) is from the breaking 
$SU(2)_R\times U(1)_{B-L}\rightarrow U(1)_Y$ caused by the VEV $\VEV{C}$.
Since the gauge couplings at the SUSY breaking scale $M_{SB}$
are given by
\bequ
\alpha_i^{-1}(M_{SB})=\alpha_G^{-1}(\Lambda_G)
+\frac{1}{2\pi}\left(b_i\ln\left(\frac{\Lambda_G}{M_{SB}}
               \right)\right)\quad,\quad(i=1, 2, 3)
\eequ
where $\alpha_G^{-1}(\Lambda_G)\sim 25$ and 
$\Lambda_G\sim 2\times 10^{16}$ GeV, 
the unification conditions $\alpha_1(\Lambda_A)=\alpha_2(\Lambda_A)$,
$\alpha_1(\Lambda_A)=\alpha_3(\Lambda_A)$ and
$\alpha_2(\Lambda_A)=\alpha_3(\Lambda_A)$ can be rewritten 
\begin{eqnarray}
&&\left(\frac{\Lambda_A}{\Lambda_G}\right)^{14}
\left(\frac{\Lambda_C}{\Lambda_A}\right)^6
\left(\frac{\det \bar M_L}{\det \bar M_{D^c}}\right)
\left(\frac{\det \bar M_Q}{\det \bar M_{U}}\right)^4
\left(\frac{\det \bar M_Q}{\det \bar M_{E^c}}\right)^3
\left(\frac{\det \bar M_W}{\det \bar M_{X}}\right)^5  \nn \\ 
&&\qquad 
=\Lambda_A^{-\bar r_{D^c}+\bar r_L-4\bar r_{U^c}-3\bar r_{E^c}+7\bar r_Q
-5\bar r_X+5\bar r_W}, \\
&&\left(\frac{\Lambda_A}{\Lambda_G}\right)^{16}
\left(\frac{\Lambda_C}{\Lambda_A}\right)^4
\left(\frac{\det \bar M_{D^c}}{\det \bar M_{L}}\right)
\left(\frac{\det \bar M_Q}{\det \bar M_{U}}\right)
\left(\frac{\det \bar M_Q}{\det \bar M_{E^c}}\right)^2
\left(\frac{\det \bar M_G}{\det \bar M_{X}}\right)^5  \nn \\ 
&&\qquad =
\Lambda_A^{-\bar r_{L}+\bar r_{D^c}-\bar r_{U^c}-2\bar r_{E^c}+3\bar r_Q
-5\bar r_X+5\bar r_G}, \\
&&\left(\frac{\Lambda_A}{\Lambda_G}\right)^{4}
\left(\frac{\det \bar M_{D^c}}{\det \bar M_{L}}\right)
\left(\frac{\det \bar M_U}{\det \bar M_{Q}}\right)
\left(\frac{\det \bar M_G}{\det \bar M_{W}}\right)^2
\left(\frac{\det \bar M_G}{\det \bar M_{X}}\right) \nn \\ 
&&\qquad =\Lambda_A^{-\bar r_{L}+\bar r_{D^c}-\bar r_{Q}+\bar r_{U}-2\bar r_W
-\bar r_X+3\bar r_G}.
\end{eqnarray}
Here 
$\bar r_I$ are the ranks of the mass matrices of superheavy fields 
$\bar M_I$. 
Note that the above conditions are dependent only on the ratio of the
determinants of mass matrices which are included in the same multiplet 
of $SU(5)$ and on the symmetry breaking scales $\Lambda_A$, $\Lambda_C$.
If all the component fields in a multiplet have been superheavy, the
above ratios would be of order one,  because the determinants are given
by $\det \bar M =\lambda^{\sum_i c_i}$. However, since part of the component
fields (massless Higgs doublets or Nambu-Goldstone modes) do not appear 
in the mass matrices, the above ratios are dependent only 
on the charges of these massless modes.
If all the other fields than in MSSM become
superheavy, the above ratios are easily estimated as
\begin{eqnarray}
\frac{\det \bar M_L}{\det \bar M_{D^c}}&\sim & \lambda^{-2h} \\
\frac{\det \bar M_Q}{\det \bar M_{E^c}}\sim 
\frac{\det \bar M_{U^c}}{\det \bar M_{E^c}}&\sim &\lambda^{c+\bar c-2a} \\
\frac{\det \bar M_G}{\det \bar M_X}\sim 
\frac{\det \bar M_W}{\det \bar M_X}&\sim &\lambda^{-2a}.
\end{eqnarray}
Then the conditions for the coupling unification becomes
\begin{equation}
\Lambda \sim  \lambda^{\frac{h}{7}}\Lambda_G,\label{cond1},\ 
\Lambda \sim \lambda^{-\frac{h}{8}}\Lambda_G,\label{cond2},\ 
\Lambda \sim \lambda^{-\frac{h}{2}}\Lambda_G.\label{cond3}
\end{equation}
So the unification conditions become $h\sim 0$, and thus 
the cutoff scale must be taken as $\Lambda\sim \Lambda_G$.
It is obvious that if the cutoff scale have been another scale (for example,
the Planck scale), in MSSM three gauge couplings would meet at the scale.
This means that in this scenario it is not accidental that three gauge 
couplings meet at a scale in MSSM, even though the unification scale in
our scenario is different from the usual unification scale.
Note that the above results are independent of the detail of the Higgs 
sector, because the requirement that all the other fields than those in MSSM
become superheavy determines the field content of the massless fields,
whose charges are important to examine whether gauge couplings meet
at the unification scale $\Lambda_A$ or not. The above argument can 
be applied also to the scenario of $E_6$ unification, though instead
of usual doublet Higgs charge $h$ we have
to use effective Higgs charges
$h_{eff}\equiv h+\frac{1}{4}(\phi-\bar \phi)$,
where $E_6$ is broken into $SO(10)$ by the VEV 
$|\VEV{\Phi}|=|\VEV{\bar \Phi}|\sim \lambda^{-\frac{1}{2}(\phi+\bar\phi)}$.

Note that the condition $h\sim 0$ does not mean $h=0$, 
because there is an ambiguity involving
order 1 coefficients and we have used only one loop RGEs. 
However, the above analysis
is fairly useful to provide a rough picture of the behavior. 

\section{Proton decay}
The proton decay via dimension 5 operators\cite{SY} is suppressed
in our scenario, because effective colored Higgs mass is given by
$m_H^c\sim \lambda^{2h}$ and the dimension 5 operators are suppressed
due to anomalous $U(1)_A$ gauge symmetry. On the other hand, proton decay
via dimension 6 operators is reachable, because the cutoff scale must be
around the usual GUT scale $\Lambda_G\sim 2\times 10^{16}$ GeV and the 
unification scale is given by $\lambda^{-a}$. If we adopt $a=-1$, then
the lifetime of the proton is roughly estimated as
\begin{equation}
\tau_p(p\rightarrow e\pi^0)\sim 3\times 10^{33}
\left(\frac{\Lambda_A}{5\times 10^{15}\ {\rm GeV}}\right)^4
\left(\frac{0.015({\rm GeV})^3}{\alpha}\right)^2  {\rm years},
\end{equation}
which is near the present experimental lower bound.\cite{SKproton}.
Here $\alpha$ is the hadron matrix element parameter. Here we use the
formula\cite{hisano} and the value $\alpha$ given by lattice calculation.
\cite{lattice}
Though the prediction is strongly dependent on the actual unification
scale which is dependent on the order one coefficients, this rough estimation
gives a strong motivation for the future experiments for proton decay search.

\end{document}